\documentclass{article}

\usepackage{arxiv}

\usepackage[utf8]{inputenc} 
\usepackage[T1]{fontenc}    
\usepackage{hyperref}       
\usepackage{url}            
\usepackage{booktabs}       
\usepackage{amsfonts}       
\usepackage{nicefrac}       
\usepackage{microtype}      
\usepackage{lipsum}		
\usepackage{graphicx}
\usepackage{natbib}
\usepackage{doi}
\usepackage{amsmath}
\usepackage{relsize}
\usepackage{exscale}
\usepackage{graphics}
\usepackage{mathtools}
\usepackage{amsthm}
\usepackage{longtable}
\usepackage[english]{babel}
\usepackage{graphicx}
\usepackage{bmpsize}
\usepackage{soul}
\usepackage{color}
\usepackage{subfiles}
\usepackage{tabularx}

\title{Modelling fine-sliced three dimensional electron diffraction data with dynamical Bloch-wave simulations}

\author{ \href{}{\hspace{1mm}Anton Cleverley} \\
	Department of Chemistry\\
	University of Warwick\\
	Coventry, CV4 7AL \\
	\texttt{} \\
	\And
	\href{}{\hspace{1mm}Prof. Richard Beanland} \\
	Department of Physics\\
	University of Warwick\\
	Coventry, CV4 7AL \\
	\texttt{r.beanland@warwick.ac.uk} \\
}

\date{}

\renewcommand{\headeright}{}
\renewcommand{\shorttitle}{Modelling fine-sliced three dimensional electron diffraction data with dynamical Bloch-wave simulations}

\hypersetup{
pdftitle={A template for the arxiv style},
pdfsubject={q-bio.NC, q-bio.QM},
pdfauthor={},
pdfkeywords={First keyword, Second keyword, More},
}

\begin{document}
\maketitle

\begin{abstract}
\noindent Recent interest in structure solution and refinement using electron diffraction (ED) has been fuelled by its inherent advantages when applied to crystals of sub-micron size, as well as a better sensitivity to light elements.  Currently, data is often processed using software written for X-ray diffraction, using the kinematic theory of diffraction to generate model intensities -- despite the inherent differences in diffraction processes in ED.  Here, we use dynamical Bloch-wave simulations to model continuous rotation electron diffraction data, collected with a fine angular resolution (crystal orientations of $\sim0.1^\circ$).  This fine-sliced data allows us to reexamine the corrections applied to ED data.  We propose a new method for optimising crystal orientation, and take into account the angular range of the incident beam and varying slew rate.  We extract observed integrated intensities and perform accurate comparisons with simulations using rocking curves for a (110) lamella of silicon 185~nm in thickness.  $R_1$ is reduced from $26\%$ with the kinematic model to $6.8\%$ using dynamical simulations.  
\end{abstract}
\keywords{Electron Diffraction \and Structure Solution \and Bloch-Waves}
\section{Introduction}
\label{intro}
Electron diffraction (ED) is currently enjoying increased attention and activity due to its ability to work with crystallites that are far smaller than can be tackled by X-ray diffraction (XRD) \citet{Xu2019a}.  Structural solution utilising ED has dramatically increased since the turn of the century due to advances in computer control and detector development \citet{Gemmi19} and the new methodologies that have been developed for structure solution are generally known by the term three-dimensional electron diffraction (3D-ED) \citet{Gemmi1}.  Just as in XRD, these techniques measure the direction and intensity of many Bragg-diffracted beams from a crystal, which are then processed to deduce a unit cell, given Miller indices $hkl$ and observed intensities $I_{hkl}^{(obs)}$.  These data can then be used to produce a crystal model using structure solution methods.  Currently, many analyses of ED data for structural solution and refinement use well-established and relatively sophisticated XRD software, despite the vastly different scattering processes involved.  As a result, the quality of structural solutions from 3D-ED appears much worse than that of XRD, even though the structures obtained seem reliable \citet{Palatinus2015a}.  To develop the field of ED further, it is necessary to improve quality of fit.

Electron detector technology has seen a significant improvement in both quantum efficiency and speed in recent years \citet{FARUQI2018180} \citet{PATERSON2020112917}, allowing ever greater amounts of data to be obtained.  The arrival of fast pixelated detectors is fuelling a trend away from integrated intensities and it is now possible to collect data that has quite fine resolution both temporally and in scattering angle.  The many differences between electrons and X-rays when used for structure solution are well documented \citet{Gemmi:lu5002}, but can seem relatively subtle in integrated data. Conversely, in fine-sliced data these differences can be observed more clearly and there is sufficient information to allow them to be modelled more comprehensively.  In this work, we explore continuous-rotation electron diffraction (cRED) data taken with a crystal orientation resolution of $\sim0.1^{\circ}$, in combination with Bloch-wave electron diffraction simulations.  Our aim is to elucidate the most important experimental and modelling parameters that will be necessary in future dynamical refinement methodologies.

To a great extent ED and XRD are complementary, with strengths in different areas that can be very powerful when used in combination \citet{Yun:ro5003}. There are many differences between ED and XRD, including very different wavelengths and damage mechanisms, but the principal one which affects diffracted intensities is the strength of interaction, with electrons roughly 10,000 times more likely to be scattered than X-rays \citet{Xu2019a}.  Thus, multiple scattering is usual for electrons -- and is essential to capture the interaction of a fast electron with even a single gold atom \citet{Howie2014} -- whereas single scattering usually dominates for XRD. Since structural refinement relies on minimising the difference between $I_{hkl}^{(obs)}$ and model calculated values $I_{hkl}^{(calc)}$, it is therefore unsurprising that a fit to a single (kinematic) scattering model is poor for a method where multiple (dynamical) scattering dominates.  

Apart from the real difference in the current state of development of dedicated software for ED and XRD data analysis, the main reason for the continued adherence to a scattering model that is known to be inadequate for ED is the relative difficulty of calculation for dynamical scattering in comparison with the kinematic model.  In both models, the starting point for calculation of the diffracted intensity for reflection $\mathbf{g} = hkl$ is the structure factor $F_{hkl}$
\begin{equation}
  F_{hkl} = \sum_{j=1}^{N}f_j(\theta_B) T_j \exp \left(2 \pi i \mathbf{g} \cdot \mathbf{r}_j \right), \label{StrucFac}
\end{equation}
where $f_j(\theta_B)$ is the atomic scattering factor evaluated at the Bragg angle $\theta_B$, $T_j$ the thermal factor and $\mathbf{r}_j$ fractional atomic coordinates of the $j^{th}$ atom, the sum taken over all $N$ atoms in the unit cell.  In the kinematic model, it is commonly stated that the structure factor $F_{hkl}$ gives the complex amplitude of the diffracted beam, whose intensity is proportional to $I_{hkl}^{(kin)} = |F_{hkl}|^2 = F_{hkl} F^*_{hkl}$, where * indicates complex conjugate.  With tabulated scattering factors Eq.~(\ref{StrucFac}) can be evaluated almost instantaneously on even the most basic computer. In comparison, modelling dynamical scattering for ED to obtain $I_{hkl}^{(dyn)}$ requires solving Schrodinger's wave equation for an electron travelling through the crystal, usually done using either the Bloch-wave method or a multislice wave scattering/propagation calculation.  The structure factor enters the Bloch wave calculation as elements in the scattering matrix, which contains all excitetd $\mathbf{g}$-vectors and their differences, but is present only indirectly in a multislice calculation. These models, which require significant computing resources and may be cluster or GPU-based, are widely used in more traditional electron diffraction work such as convergent beam electron diffraction (CBED) and in the simulation of transmission electron microscopy (TEM) and scanning TEM (STEM) images.  Dynamical modelling has been applied to ED \citet{DUDKA2007474} \citet{SinklerMarks+2010+47+55} \citet{Palatinus:td5023}, but to date has only been incorporated into dedicated 3D-ED analysis in the software PETS \citet{palat_2011}.

\begin{figure}
    \caption{An illustration of the range of intensities that are obtained for the kinematically-forbidden $\bar{14}~\bar{2}~\bar{2}$ reflection in silicon with a thickness of 180nm in a dynamical simulation.  The image below shows the corresponding LACBED pattern (see section \ref{Si})}
    \label{1422}
    \centering \includegraphics[width=1.0\columnwidth]{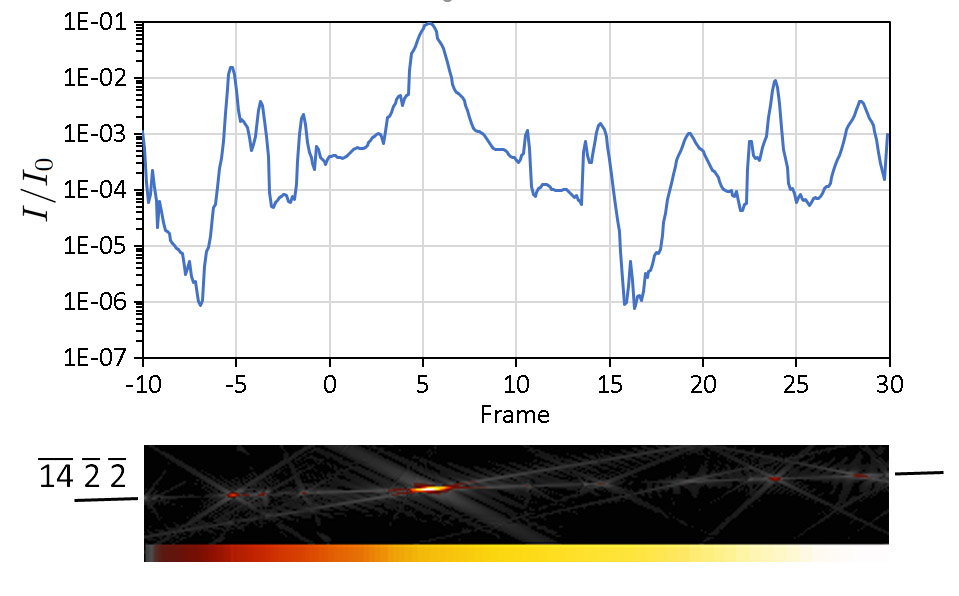}
\end{figure}

In any quantitative experiment, measurements and calculations must meet at some point and in XRD it is most convenient for that point to be the structure factor.  The correspondence between model and experiment is usually captured using an $R$-factor (see \ref{Rfac_eqns}).  Although it is often said that diffracted X-ray intensities from a crystal are given  by $|F_{hkl}|^2$ (Eq.\ref{StrucFac}), in reality things are not so simple and many other factors need to be accounted for, which depend on both the experiment being performed and the material.  These factors are usually considered to be experimental, and are dealt with and refined separately from the crystal structure itself. In Eq.~\ref{Corr} we include these other factors explicitly and make a distinction between experimental measurements of diffracted intensity $I_{hkl}^{(expt)}$ and the `observed' intensities $I_{hkl}^{(obs)}$ that are suitable for comparison  with calculations $|F_{hkl}^{(calc)}|^2$:
\begin{equation}
  |F_{hkl}^{(obs)}|^2 = I_{hkl}^{(obs)} = L^{-1} G^{-1} B^{-1} A^{-1} E^{-1} S^{-1} C^{-1} M^{-1} p^{-1} I_{hkl}^{(expt)}, \label{Corr}
\end{equation}
with corrections for Lorentz factors $L$,  geometry $G$, background $B$, absorption $A$, extinction $E$, scaling $S$, fluctuations in the incident X-ray beam intensity $C$, mosaicity $M$ and polarisation $p$.  It is usually assumed that these corrections are independent and commutative, i.e. can be applied in any order, although they can be refined iteratively \citet{ladd_palmer_1994}.  In XRD each reflection $hkl$ has a single well-defined $I_{hkl}^{(obs)}$, so that if it is sampled multiple times, or there are symmetrically equivalent reflections, they can all be merged into a single measurement with improved fidelity.  Comparison with theory is then made using a metric $R_{merge}$.  However, in dynamical diffraction a reflection no longer has a single well-defined intensity \citet{Clabbers:td5055}, as illustrated by Fig.~\ref{1422}.  This shows a Bloch-wave simulation of a $\bar{14}~\bar{2}~\bar{2}$ reflection from the silicon cRED data set (section \ref{Si}) which, although kinematically forbidden, has intensities up to 10\% of the incident beam intensity where pathways for multiple allowed reflections exist.  Although this is an extreme example, it is not uncommon for weak reflections to be affected in this way in ED.  Thus an average intensity, taken either by merging multiple measurements, symmetrically equivalent reflections, or through the use of a precessed incident beam \citet{OLEYNIKOV2007523}, will in general converge to some ill-defined value. Conversely, comparison of experimental data with a dynamical model in which each measurement is simulated individually should give a better $R_1$ without any data merging \citet{Palatinus2015a}. 

Before embarking on a structural refinement in which $r_j$ and $T_j$ in Eq. \ref{StrucFac} are determined for each atom in the unit cell by minimising $R_1$, the many corrections in Eq. \ref{Corr} must be applied to optimise the experimental input $|F_{hkl}^{(obs)}|$.  At first sight, it might therefore appear that Eq.~\ref{StrucFac} deals with scattering theory while Eq.~\ref{Corr} deals only with experimental parameters, but this is not strictly correct -- absorption and extinction, for example, are scattering effects that depend upon the sample.    For dynamical electron diffraction, we must reconsider the validity of Eq. 2 to account for the differences in ED vs XRD experiments.  For electrons, we can discard the polarisation correction $p$ since electron beams are unpolarised, but each of the other terms has an equivalent in ED. Thus, before presenting our results from a continuous rotation electron diffraction (cRED) measurement, we briefly discuss each in turn.

i) \textit{Lorentz corrections} $L$.  Lorentz corrections ensure that a given reflection has the same integrated intensity in XRD irrespective of the way the crystal is rotated, i.e. they account for the different time spent in the vicinity of each Bragg condition during data collection.  As purely geometrical corrections, they apply equally to XRD and ED \citet{Palatinus:td5023, ZhangOHZ} although if beam precession is employed this must also be taken into account \citet{GJONNES19971, ZHANG201047}.  In our cRED experiment, we measure the integrated intensity of each reflection individually from rocking curves in both experiment and simulation and it is therefore possible to compare intensities without applying Lorentz corrections.  However, to maintain equivalence with XRD refinement methods it is preferable to apply them, and we do so for both experiment and simulation.

ii)	\textit{Geometry} $G$.  Corrections for other geometrical issues, such as variations in specimen height or rotation axis, are needed in XRD and also for ED.  These issues primarily affect a diffracted beam's position, rather than the intensity that is our main interest here.  It is of course important to ensure that the crystal of interest does not wander out of the electron beam as it is rotated \citet{Cichocka:yr5038, PLANARUIZ2020112951}.  More importantly, since dynamical diffraction is exquisitely sensitive to geometry, crystal orientation must be known to high precision. At crystal orientations where diffraction is strong, a good calculation of intensities in a cRED measurement using dynamical scattering requires an angular precision and accuracy better than 40 arc seconds ($\sim$0.2 mrad). Previously, orientation refinement has been performed by optimising the fit between experimental and dynamical integrated intensities \citet{Palatinus2013}.  Here, we show that fine-sliced data allows a quicker and more straightforward orientation refinement to be performed, using the sequence of reflections as they appear during crystal rotation. This also permits a measurement of varying slew rate (section \ref{Si}). As a corollary of the requirement for high precision in crystal orientation, the angular range of the incident beam must also be taken into account; while parallel illumination is often assumed, in practice there is always some beam convergence or divergence that broadens the range of reciprocal space that is sampled.

iii) \textit{Background} $B$.  In ED, corrections for background need to be considered from two main sources, a) the detector; and b) non-Bragg scattered electrons (both elastic and inelastic). For (a), detector characteristics such as dark noise levels, quantum efficiency and linearity can be measured effectively; these corrections are necessary but straightforward.  Unfortunately, (b) is rather more problematic.  Amorphous material, e.g. the support film for the crystal, can create a non--linear background \citet{tivol_2010}, but the sample itself also produces non-Bragg thermal diffuse scattering (TDS, caused by displacement of atoms from their mean positions by thermal vibrations) and inelastic scattering. Importantly, these electrons can be diffracted again by the crystal, producing a background that is highly structured, with Kikuchi lines and dynamical scattering effects \citet{EGGEMAN20121}. Calculating this background is not a trivial exercise and generally requires a model of thermal vibrations (ideally, complete knowledge of the phonon spectrum) \citet{ AMULLER2001371, Kolb2012UnitingEC} and a full quantum-mechanical description of inelastic processes \citet{FORBES20111670} respectively.  In a complete model of electron scattering these effects would be taken into account in the calculation of diffracted intensities to be compared against experiment.  Currently, while some current multislice simulation packages can do so \citet{OleynikoveMap} none have yet been implemented for 3D-ED experiments.  Here, we use a simple Bloch-wave model that neglects this `background' intensity of diffuse scattering by the crystal.

iv) \textit{Absorption} $A$. This is another term which, strictly speaking, should be considered in scattering theory but in XRD its behaviour is simple enough for it to be corrected as an experimental variable.  For the energies typical of ED (80-300 keV) and a thin specimen suitable for structure solution, true absorption of the electron beam does not happen to any appreciable extent.  However, the attenuation of a Bragg reflection, as electrons are scattered into the diffuse background by TDS or inelastic interactions, is a very similar effect.  In ED, TDS is enhanced significantly when the electrons are channeled along atom columns, particularly those with high atomic number \citet{TDS_1965}.  Thus, dark bands can be seen between low-index Bragg conditions in bright field LACBED patterns (`anomalous absorption', \citet{HHNPW, JORDAN1991237} see examples below).  This complicated behaviour means that in ED it is best dealt with in scattering theory, and should no longer regarded as an experimental parameter. Both Bloch-wave and multislice models can account for this effect in ED.

v) \textit{Extinction} $E$. This is simply the X-ray term for dynamical diffraction effects.  The underlying theory is very similar; the two-beam analysis by Darwin \citet{darwin1914a, darwin1914b, darwin1922} for X-rays has many resemblances to that of Howie and Whelan for electrons \citet{HowWhel1961, HHNPW}, so much so that they are sometimes referred to as the Darwin-Howie-Whelan model \citet{james1990}. The term \textit{extinction} refers to the transfer of intensity from the direct beam to a diffracted beam $\mathbf{g}$, and back again, as a function of crystal thickness; the distance over which this occurs is known as an extinction distance $\xi_g$. Like absorption, in XRD this can be considered an experimental correction, since X-ray extinction distances are usually much larger than the size of a crystallite.  In ED, where extinction distances can be tens of nm, this is not the case. (In XRD, there is also `secondary' extinction, which refers to the enhanced absorption of strong diffracted beams in large crystals \citet{ladd_palmer_1994}, whose counterpart in ED is anomalous absorption, above). Again, this correction should not be applied to ED data, but taken into account in the calculation of diffracted intensities.

vi) \textit{Scaling} $S$.  This correction takes account of the varying proportion of the incident beam occupied by a crystal of irregular shape as it is rotated.  This is certainly a correction that should be applied in principle in ED, although in practice extreme care must be taken not to confound it with simple loss of intensity in the direct beam due to a very strong diffracted beam, or the effect of `absorption' due to TDS.  ED has a potential advantage over XRD here, in that a crystal can be imaged directly, allowing scaling to be calculated from a series of images taken after a diffraction measurement \citet{PLANARUIZ2020112951}.  Without these images, scaling of intensities is difficult to calculate, although here we propose a method using the direct beam intensity.
 
vii) \textit{Fluctuations in incident electron beam intensity} $C$.  cRED measurements are in general very rapid; diffracted electron beam intensities are sufficiently high that they can be sampled to good precision even in a small fraction of a second and the total time for data collection is often less than a minute.  Variations in incident beam intensity on this timescale are negligible.  Despite this, it is common to see significant changes in the direct 000 beam intensity in a cRED dataset (e.g. SI video 1).  This happens because electron diffraction is strong and the crystal occupies much, or all, of the incident electron beam. In ED, it is quite possible for a diffracted beam with a large structure factor to have a higher intensity than the direct beam.  Nevertheless, this effect is not due to a change in incident beam intensity and therefore this correction is not appropriate for a cRED measurement.
 
vii) \textit{Mosaicity} $M$. Crystal imperfections, in the form of dislocations, low-angle grain boundaries and cracks, allows the Bragg condition to be satisfied over a wider range of angles than would be the case for a perfect crystal without strain.  In XRD, the presence of these defects can be helpful in that they effectively break the crystal into a mosaic of small crystal blocks that reduces extinction effects significantly, \citet{ladd_palmer_1994} although it can also produce broadening of diffracted beams.  For high energy electrons, with much smaller extinction distances, defects alter diffracted intensity very strongly, allowing them to be visualised directly in diffraction-contrast TEM \citet{HHNPW, WillCart}.  In ED they may have a significant effect on measured intensities that is too complicated to account for in any simple model and will vary from one crystal to another in an unknown way.  Currently, their effect is neglected completely and this is probably the best approach until the more tractable effects of dynamical diffraction are fully accounted for.

To summarise, in reconsidering Eq. \ref{Corr} for dynamical ED structure solution or refinement, we find that some factors that can be separated from the scattering model and considered `experimental' variables for XRD -- i.e. absorption $A$, geometry $G$, and extinction $E$ -- must instead be considered explicitly in the scattering model for ED.  Other experimental factors -- polarisation $p$, incident beam intensity variations $C$, and the complicated effects of mosaicity $M$ -- can probably be neglected at the current level of simulation fidelity, being either relatively unimportant or too difficult to tackle with current methods.  Finally, truly experimental corrections $L$, $B$ and $S$ that both XRD and ED hold in common must be taken into account, but the differences in Bragg angle and hardware means that they are rather different for ED.  One of the biggest changes in emphasis is that the point of contact between experimental measurements and modelled intensities in a dynamical refinement no longer has the simple and elegant interpretation relating to the structure factor.

In what follows, we still use the square root of the corrected intensity as a metric for $R_1$, and use the same notation, i.e. $F_{hkl}^{(obs)}$, bearing in mind that the relationship between these observed values and the structure factor $F_{hkl}$ of Eq.~\ref{StrucFac} is no longer straightforward.  Our discussion above justifies only corrections for Lorentz factors $L$, background $B$ and scaling $S$:

\begin{equation}
  |F_{hkl}^{(obs)}|^2 = I_{hkl}^{(obs)} = B^{-1} L^{-1} S^{-1} I_{hkl}^{(expt)}. \label{CorrDyn}
\end{equation}

We compare $F_{hkl}^{(obs)}$ with the square root of calculated ED integrated intensities using a dynamical diffraction model in which extinction $X$ and absorption $A$ are implicit, resulting from the crystal itself. Dynamical integrated intensities $I_{hkl}^{(dyn)}$ are a non-linear function of crystal thickness $t$, geometry $G$ and beam profile $P$, also with corrections for Lorentz factors $L$:

\begin{equation}
  |F_{hkl}^{(dyn)}|^2 = L^{-1}~I_{hkl}^{(dyn)}\left( X~t~G~P~\right), \label{ImpDyn}
\end{equation}

We now explore this approach using Bloch-wave simulations to compare kinematic and dynamical model fits to experimental data for simple, well-known materials.  We examine data from a single crystal of silicon, ion milled to a thin foil.  This almost perfect crystal shows strong dynamical scattering and is used to evaluate the improvement in fit using dynamical modelling as well as the importance of correction factors applied to the raw data.

\section{Experiment}

\subsection{Data collection}
Experimental data was obtained using selected area electron diffraction (SAED) with parallel beam illumination on a JEOL 2100 LaB$_6$ transmission electron microscope operating at 200 kV. The sample was a defect-free single crystal ion milled to produce a (110) lamella with extensive electron transparent regions.  Diffraction patterns were produced using a strongly excitetd third condenser lens giving close to parallel illumination and a selected area aperture, and captured using a Gatan OneView camera recording continuously at 8$\times$ binning ($4096\times4096 \rightarrow 512\times512$) at 75 frames$\slash$sec.  (N.B. with 8x binning for data collection, the point spread function of the camera is negligible.)  The crystal was much larger than the selected area aperture and was rotated at maximum slew rate about the $\alpha$-tilt axis over $143^\circ$, giving 1389 frames, each with a nominal crystal rotation of $0.1035^\circ$ (i.e. all data collected in 18.5 seconds).

\subsection{Data reduction and dynamical simulations}
$PETS$ (Process Electron Tilt Series) is a software program developed specifically to process electron diffraction tilt series data \citet{palat_2011}. We used $PETS$ to find the crystal orientation and unit cell, to index reflections and give their rocking curves, and obtain a $hkl$ and $I_{hkl}^{(obs)}$ list as detailed in sections 4.1-4.6 of \citet{Palatinus15}.

Bloch wave simulations were performed using the code $Felix$ \citet{felix} running on a high performance computing cluster (typically 384 cores, completing a simulation in $\sim30$ seconds) using a python script to extract data from the the dyn.cif$\_$pets file generated by PETS and write input files for $felix$, e.g. setting the incident beam orientation with respect to the crystal.  Large-angle convergent beam electron diffraction (LACBED) patterns of size 400x400 pixels were simulated over an angular range corresponding to 40 frames ($4.14^\circ$, or 72.26 mrad).  The $x$-axis of the image was taken to be along a direction perpendicular to the rotation axis.  Successive simulations overlapped by 10 frames, thus requiring 70 simulations in total to cover the full angular range of the cRED data. 

\section{Results}
\label{Si}

The Si cRED data contained 962 reflections with $I_{hkl}^{(obs)}/\sigma_{hkl} > 10$ and to avoid any issues with signal to noise the analysis is restricted to these strong reflections. The intensity of each reflection is given frame-by-frame as an output from $PETS$. Almost all these experimental rocking curves showed a single sharp peak as the crystal was rotated; only 27 had clear dynamical structure with multiple peaks.  To obtain integrated intensities according to Eq.~\ref{CorrDyn} we apply the three experimental corrections of Eq.~\ref{CorrDyn}, beginning with background correction $B^{-1}$.  As noted above, background from thermal diffuse scattering and Kikuchi lines is complex and not modelled correctly in a simple Bloch-wave simulation; here we simply subtract a linear fit to the background by interpolating beneath the peak of each rocking curve (supplementary Fig.~\ref{Backgr}).  The Lorentz correction $L^{-1}$ was then applied for each reflection $\textbf{g}$ by taking the sum of counts in the background-subtracted rocking curve and multiplying by the change in deviation parameter per frame $\delta s_g$.

\begin{figure}
    \caption{a) The relative direct beam intensity for the Si cRED data as a function of goniometer angle $\alpha$.  The red line marks the average intensity, normalised to unity.  See also Fig.~\ref{000b}.}
    \label{Si_000}
    \centering \includegraphics[width=1.0\columnwidth]{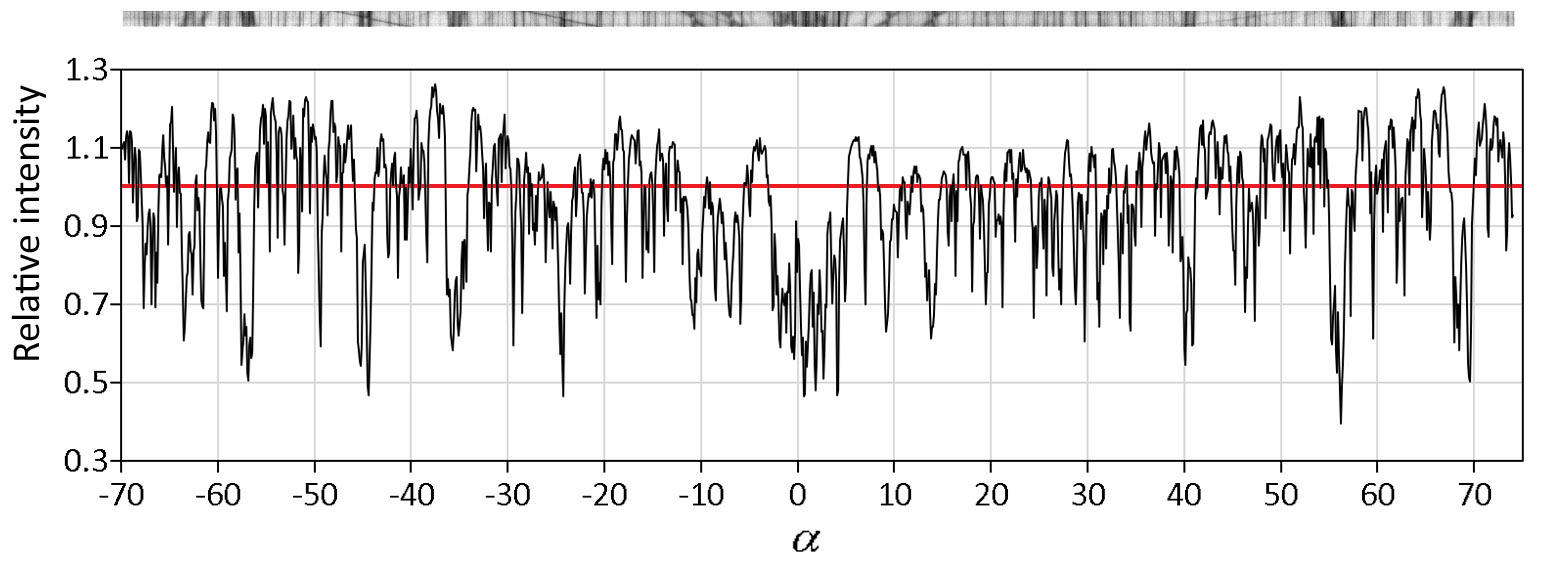}
\end{figure}

Since the silicon lamella was close to parallel-sided and much larger than the selected area aperture, it filled the field of view completely throughout data collection and therefore no frame scaling correction $S^{-1}$ should be required.  Nevertheless, it is useful to examine the intensity of the direct beam as the crystal rotated, since this should show changes in transmission that result from a varying fraction of the incident beam being intercepted by the crystal.  Thus, each frame was cropped to a $17 \times 17$ pixel image containing just the direct beam and an average beam profile $\bar{I}_{000}$ was obtained by summing all frames and dividing by $n=1387$.  A normalised direct beam intensity was then produced by dividing each cropped frame by $\bar{I}_{000}$; re-slicing this $17 \times 17 \times 1387~xyz$ data volume to $1387 \times 17 \times 17~zyx$, and taking the average along $x$, gives a $1387 \times 17$ image that shows relative direct beam intensity as the crystal is rotated (supplementary Fig.~\ref{000b}), shown in graphical form in Fig.~\ref{Si_000}.  For most of the angular range the relative intensity plot has an average with a value of unity (red line), consistent with the unchanging crystal shape sampled as it completely fills the selected area aperture.  There is a range of approx. $10^{\circ}$ at $\alpha = 0^{\circ}$ where the relative intensity is depressed by $\sim20\%$, but this is primarily due to channeling (see below); since this is calculated as part of the dynamical simulation it should not be compensated by scaling.  But most strikingly, there are many sharp and significant minima (up to $50\%$ of the relative intensity). Closer examination showed each of these dark lines to occur when a diffraction condition was satisfied; they simply indicate a transfer of intensity from the direct beam to a diffracted beam. As seen in more detail in supplementary Fig.~\ref{000b}, Bragg conditions that are satisfied in only a few frames are visible as dark vertical bands, while Bragg conditions that only pass through the direct beam slowly are also visible as inclined dark lines.

\begin{figure}
    \caption{$R_1$ calculation for cRED silicon data using the kinematic model (Eq.~\ref{StrucFac}) with $B=0.2$.   Kinematically forbidden reflections are highlighted in red.  $R_1=26.0\%$ }
    \label{kinematic_R}
    \centering \includegraphics[width=0.6\columnwidth]{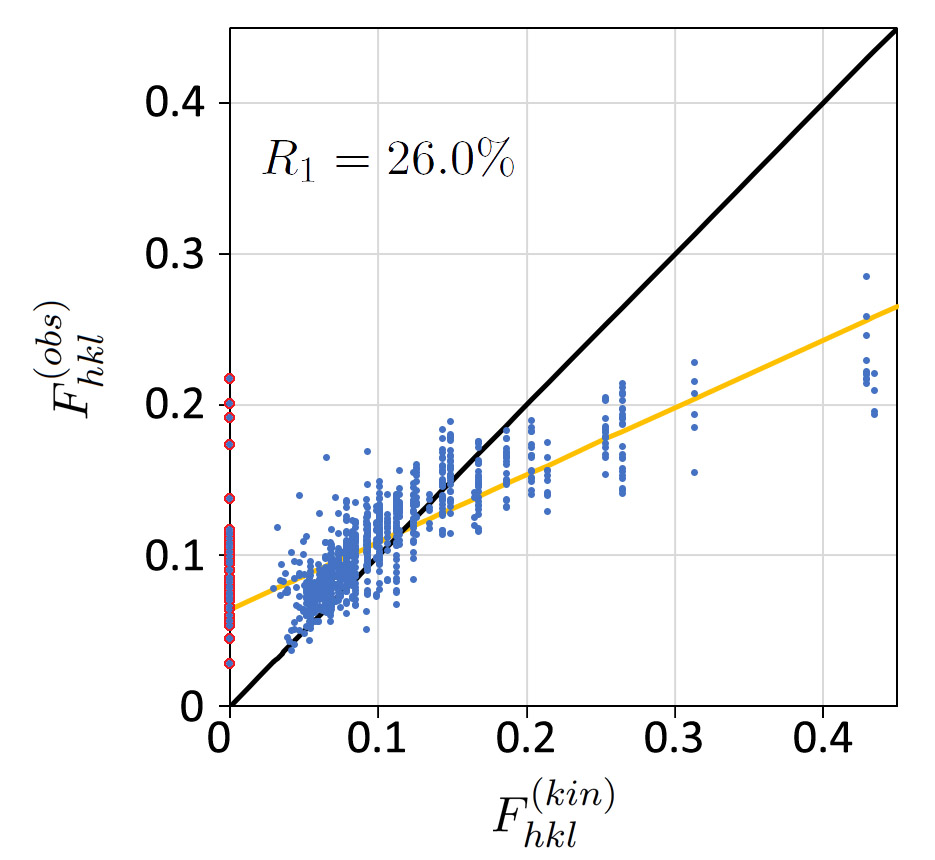}
\end{figure}

After applying corrections $B^{-1}$, $L^{-1}$ and discounting $S$ we now have a set of integrated intensities that can be compared to simulation and yield an $R$-factor. We first consider kinematic intensities calculated using Eq.~\ref{StrucFac}, taking the temperature factor to be $T_j = \text{exp}(-B_j~\text{sin}^2(\theta) / \lambda^2)$ where $B_j$ is the Debye-Waller factor.  We find a minimum $R_1=28.7\%$ at $B=0$, shown in Fig.~\ref{kinematic_R} (X-ray refinements give $B=0.54$ \citet{tobensetal}).  In this plot the bold black line indicates $R_1=0$ $\left( F^{(obs)}_{hkl} = F^{(kin)}_{hkl} \right)$ while the orange line is a least-squares linear fit to the data.  It is quite clear that the fit of the kinematic model to the data is poor, not least because there are a number of kinematically--forbidden reflections that have significant experimental intensities (highlighted in red), but also because many strong reflections ($F^{(kin)}_{hkl}>0.2$) are in fact weaker than expected .  (N.B. There are no weak reflections in the experimental data, $F^{(obs)}_{hkl}<0.04$, as we have excluded those with $I^{(obs)}_{hkl}/\sigma_{hkl} < 10$.)

\begin{figure}
    \caption{Dynamical simulation for the path of the 000 beam through reciprocal space (red line) in the silicon cRED data set, corresponding to 1387 frames (goniometer rotation of $143^\circ$ about $\alpha$).  Frame numbers are indicated; note that one division = 5 frames = $0.5175^\circ$ = 9 mrad.  Each dark line corresponds to a Bragg condition for a diffracted beam, two of which ($\bar9 \bar3 \bar3$, frames 30-50 and $311$, frames 545-590) are labelled. Below the simulation, the experimental intensity of the 000 beam is shown. Some features that are clearly present in both simulation and experiment are indicated by arrows.}
    \label{Si_track}
    \centering \includegraphics[width=0.9\columnwidth]{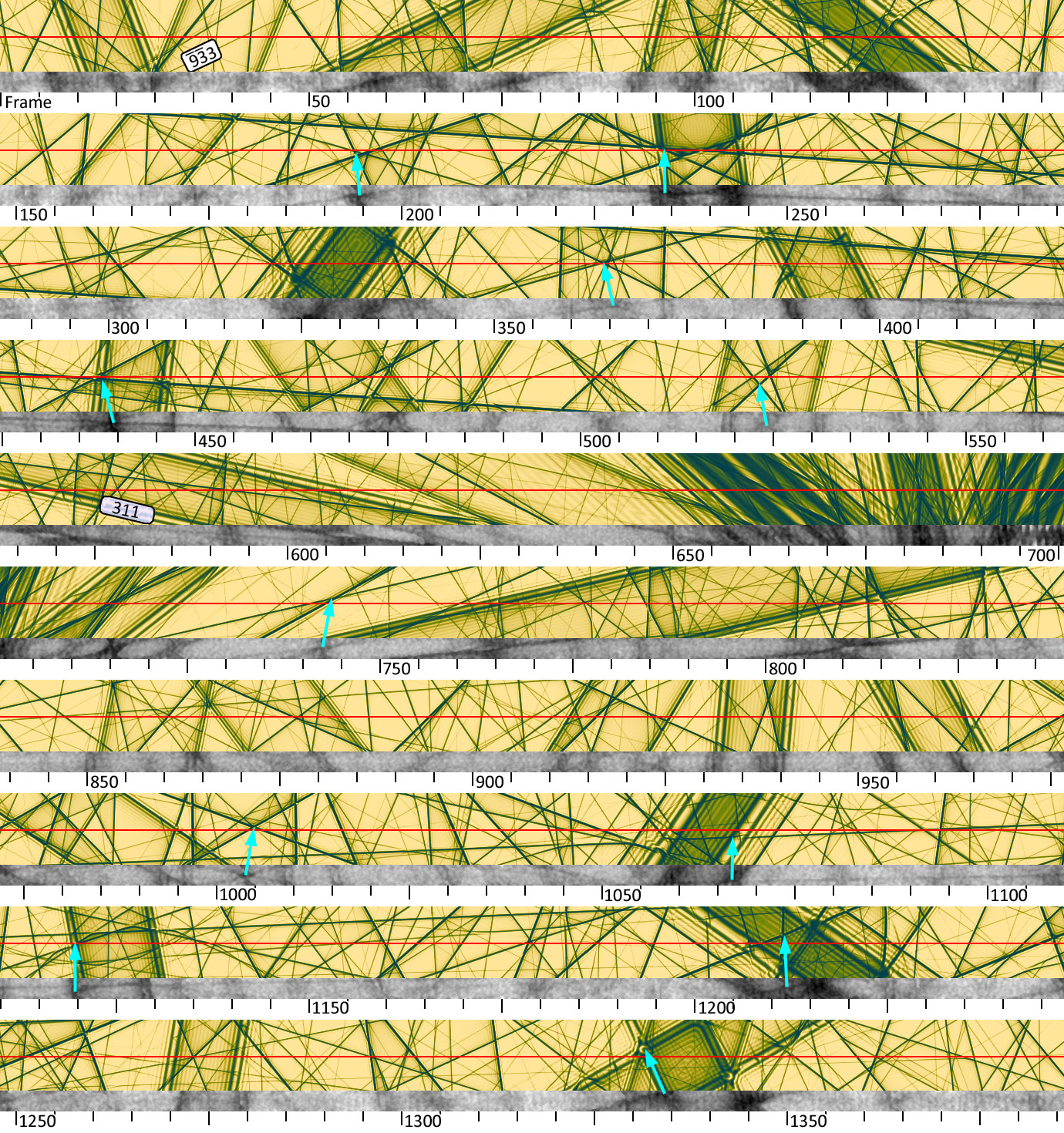}
\end{figure}

The poor $R_1$ seen in Fig.~\ref{kinematic_R} is typical of many ED refinements using a kinematic model and we now turn to a dynamical one.  We use an initial Debye-Waller factor $B=0.54$ \citet{tobensetal} since the value obtained from the kinematic model is unphysical. A Bloch-wave simulation of seventy $Felix$ 000 LACBED patterns, stitched together to make a continuous strip, is shown in Figure~\ref{Si_track}. Frame numbers underneath correspond to experiment.  In this image, a perfect plane-wave incident beam corresponds to a single point and the red line marks the nominal path traced by the direct beam through reciprocal space as the crystal is rotated.  Each dark line in the simulation shows the location of a Bragg condition; when the direct beam lies on one of these lines a diffracted beam is produced, with an intensity that can be obtained from the corresponding point in the relevant dark-field LACBED pattern.  The correspondence between this simulation and experiment can also be seen by converting the normalised experimental direct beam data volume to a 2D image, shown in greyscale below the simulation.  Some features common to both experiment and simulation are marked by arrows.

\begin{figure}
    \caption{Two examples of silicon cRED rocking curves.  (a) and (b) are experimental data. Most experimental rocking curves have a simple peak like (a) $\bar9 \bar3 \bar3$, while less than 3\% show dynamical structure like (b) 311.  c) and d) show the corresponding dark field LACBED simulations (specimen thickness 185~nm). The nominal beam path is a red line, with frame numbers in yellow.  Intensity profiles along the red line give the rocking curves (e) and (f).  The difference in frame numbers (a) to (e) and (b) to (f) are caused by a varying slew rate (Fig.~\ref{Si_corr}b). g) and h): Applying the angular spread of the incident beam (Fig.~\ref{Si_beam}) as a convolution to the simulation gives simulated rocking curves that are a good match to experiment.}
    \label{Si_rocks}
    \centering \includegraphics[width=0.6\columnwidth]{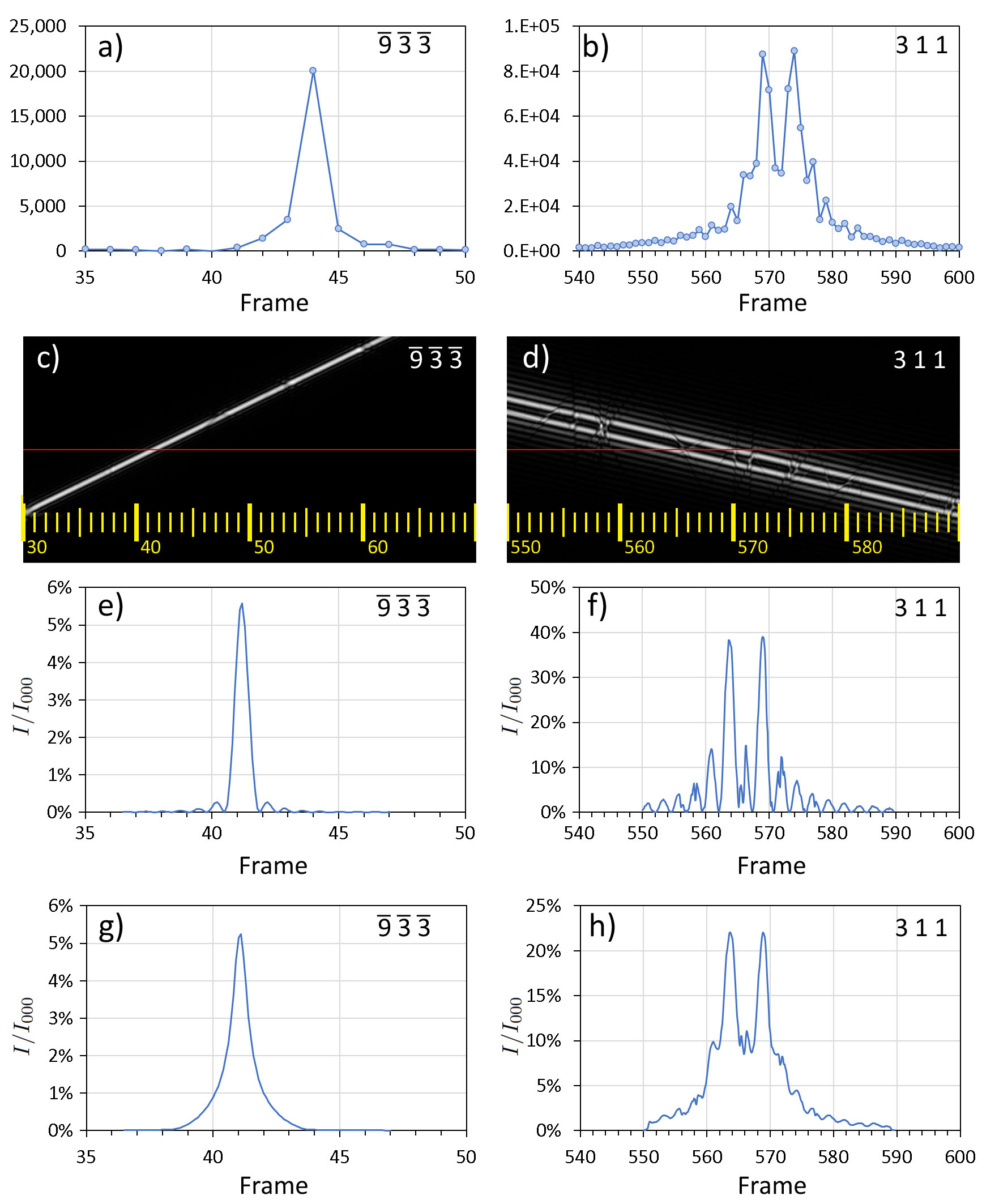}
\end{figure}

The simulated rocking curve for a reflection is given by the intensity along a line in its dark field LACBED pattern, as shown in Fig.~\ref{Si_rocks} for a) a typical reflection with a single peak $\bar9 \bar3 \bar3$, and b) one with obvious dynamical structure $311$.  Integrated intensities $I^{(dyn)}_{hkl}$ can be obtained from these simulated rocking curves in the same way as they are taken from experimental ones, although here there is no diffuse scattering and therefore no background correction.  The incident beam intensity is fixed at unity so there is no scaling correction and only Lorentz corrections need to be applied.

\begin{figure}
    \caption{$R_1$ calculation for cRED silicon data with a dynamical Bloch wave model, using the nominal direct beam path (red line in Fig.~\ref{Si_000}) and a specimen thickness of 190 nm. The black line marks perfect correspondence ($R_1=0$) and the orange line a least squares linear fit.  $R_1=12.6\%$ }
    \label{dynamic_R_0}
    \centering \includegraphics[width=0.6\columnwidth]{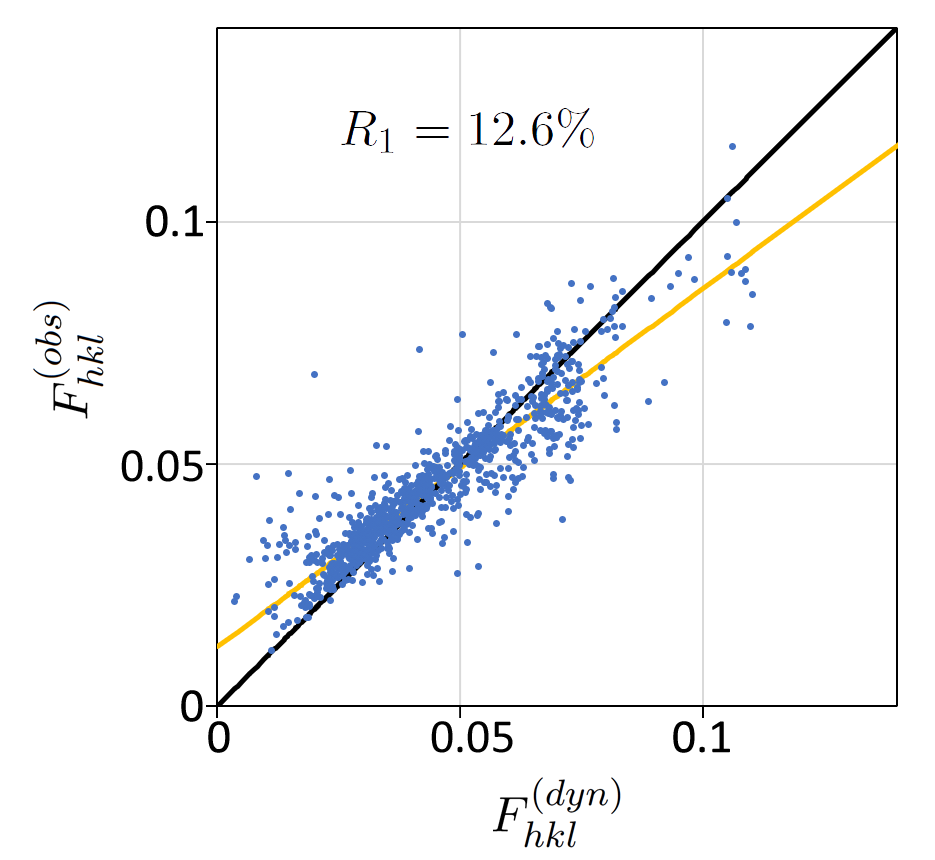}
\end{figure}

\newpage
Kinematic intensities are independent of crystal thickness, but dynamic intensities can be very sensitive to it, particularly strong reflections with short extinction distances.  It is therefore necessary to perform simulations for a range of thicknesses. (Supplementary sections \ref{Dyn_t} and  \ref{Dyn_r}). We may also expect the specimen thickness along the path of the electron beam to vary as the crystal is rotated, but this is ignored for the moment.  The best $R_1$ is obtained for a thickness of 190nm, shown in Fig.~\ref{dynamic_R_0}.  A very significant improvement over the kinematic model is apparent, with $R_1=12.6\%$.  The wide spread of intensities is no longer present, but there is still considerable scatter about the expected $R_1=0$ line and the gradient of a linear fit is 0.74.

Dynamical simulations therefore clearly give a much better fit to experimental data, as may be expected.  However, $R_1$ is still relatively high in Fig.~\ref{dynamic_R_0} and further improvements are possible by increasing the precision of the crystal orientation.  The wide range of reciprocal space covered in the simulation allows geometry to be optimised, as described below.

\subsubsection{Orientation optimisation $G$.}
\label{Si_G1}
For any given reflection in Fig.~\ref{Si_track} its Bragg condition is satisfied, and a spot will appear in the SAED pattern, when the 000 beam sits on the corresponding dark line.  The frame in which the maximum diffracted intensity appears is given by the crossing point of the Bragg condition and the red line.  It is thus possible to obtain the sequence of reflections which appear in a cRED experiment, and the frame spacing between them, for any given direct beam path.  Conversely, with knowledge of the frames in which diffracted maxima appear in an experiment we can find the corresponding path through reciprocal space.

\begin{figure}
    \caption{a) The intersections with Bragg conditions for eight reflections in the first Si $Felix$ simulation.  Frame numbers are given at the bottom of the image and the frame in which each peak intensity is found is marked by a vertical blue line. The intersection of the blue line with its Bragg condition is marked by a yellow dot (yellow lines correspond to an error of $\pm0.5$ frames).  The horizontal red line marks the nominal beam path (output from $PETS$) and if the crystal orientation was correct the yellow dots would all lie on this line. b) The best fit to a straight line is obtained by shifting the blue lines by -2.5 frames. c) $R_1$ calculation using an optimised beam path and a specimen thickness of 185 nm. $R_1=10.0\%$}
    \label{Si_intersections}
    \centering \includegraphics[width=1.0\columnwidth]{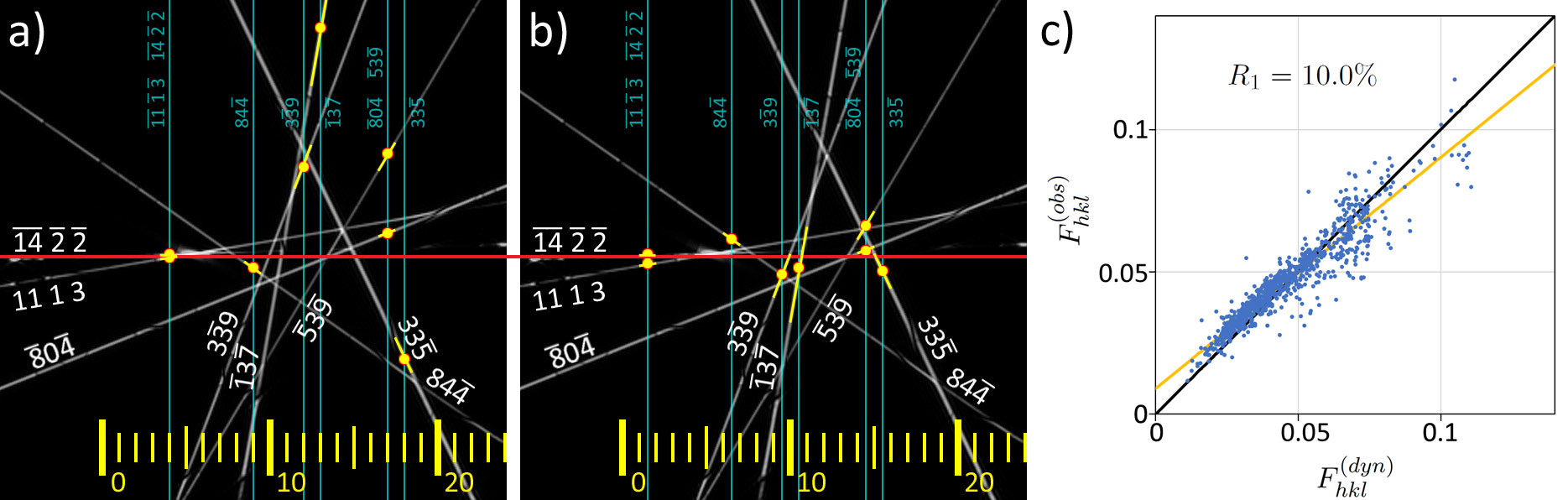}
\end{figure}

\begin{figure}
    \caption{Corrections to the nominal direct beam path (red line in Fig.~\ref{Si_track}), a) perpendicular to the line ($\beta$-tilt) and b) along the line ($\alpha$-tilt). Uncertainty in $\delta\alpha$ and $\delta\beta$ for each data point are $\sim\pm0.5$ frames, i.e. $0.05^{\circ}$. c) $R_1$ calculation after correcting integrated intensities for varying slew rate gives $R_1=9.4\%$ at a specimen thickness of 195 nm.}
    \label{Si_corr}
    \centering \includegraphics[width=1.0\columnwidth]{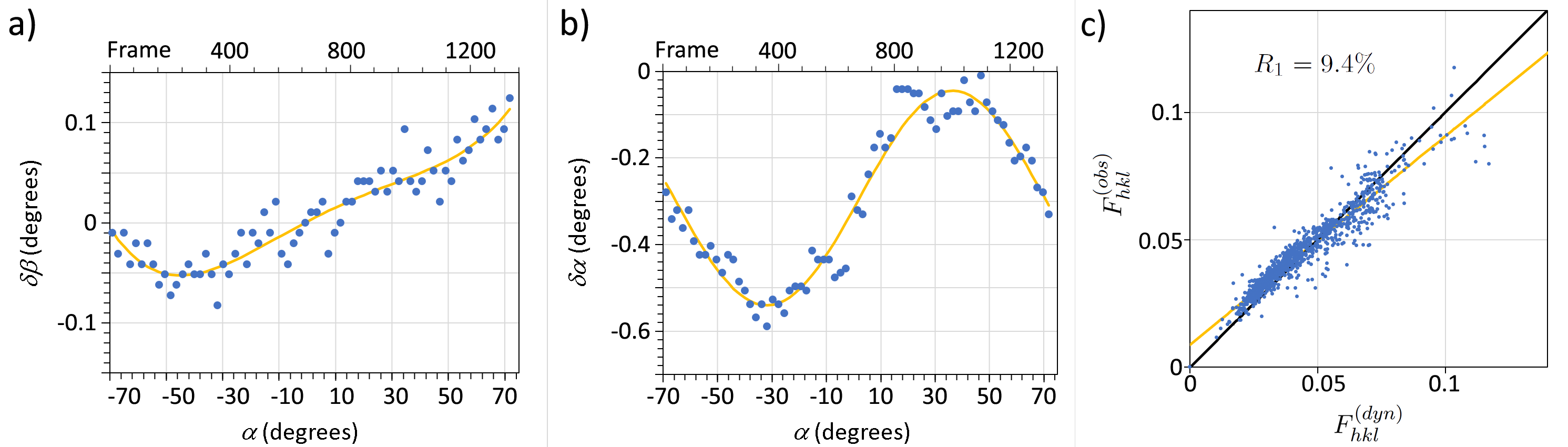}
\end{figure}

The effect of a slightly incorrect crystal orientation is shown in Fig.~\ref{Si_intersections}a.  In this simulated image each frame corresponds to a vertical stripe ten pixels wide and the nominal direct beam path runs horizontally through the centre, marked by a red line.  Experimentally, the $\bar{14}~\bar{2}~\bar{2}$ and $11~1~3$ reflections were seen in frame 4, $84\bar{4}$ was seen in frame 9, etc, as marked by blue vertical lines. The image is a superposition of eight dark-field LACBED patterns; each bright line corresponds to a diffracted beam (and to a dark line in the direct beam LACBED pattern, not shown).  The intersection of each blue line and its corresponding diffraction condition is marked by a yellow dot and must lie on the beam path; yellow lines indicate an error of 0.5 frames.  Clearly, these points do not correspond to the expected horizontal line through reciprocal space, indicating that the nominal crystal orientation is slightly in error.  An optimised crystal orientation can be found by shifting the group of blue lines while maintaining their relative frame spacing (e.g. $3\bar{3}9$ has an experimental peak intensity one frame before $13\bar{7}$ and three frames after $84\bar{4}$).  Shifting the set of blue lines by 2.5 frames to the left brings all points close to a horizontal line (Fig.~\ref{Si_intersections}b).  An optimised direct beam path was then calculated by fitting a smoothed curve to the best crystal orientation for each simulation (Fig.~\ref{Si_corr}), both for changes about the rotation axis $\delta\alpha$ and perpendicular to it $\delta\beta$ by least-squares fitting a horizontal line to the optimised set of intersection points.  Rocking curves extracted from the Bloch-wave simulations using an optimised beam path gives a significant improvement to $R_1=10.0\%$ (Fig.~\ref{Si_intersections}c), mainly by reducing the scatter in reflections with lower intensities.  This can be understood by referring back to Fig.~\ref{1422}, which shows how large variations in intensity can be found in weak beams when they coincide with stronger beams.  Optimisation of the beam path is essential to capture these interactions correctly.

The resulting corrections are shown in Fig.~\ref{Si_corr}.  The actual path traced by the direct beam deviates vertically from the red line (i.e. about the $\beta$ tilt axis) in Fig.~\ref{Si_corr} by a maximum of 10 pixels (equivalent to 1 frame, or $0.1^\circ$). There is a much larger correction needed along $\alpha$, up to 5 frames ($0.5^\circ$, Fig.~\ref{Si_corr}b) that varies through the data series, caused by a varying slew rate during rotation of the specimen. This varying slew rate is apparent in Fig.~\ref{Si_track}, where the features in the normalised experimental direct beam intensity are not found directly beneath their corresponding points in the simulation above.  A changing slew rate also has an impact on integrated intensities, since the crystal is rotating more slowly or quickly through a diffraction condition than expected.  Applying a slew rate correction to the simulated intensities gives a further improvement to $R_1=9.4\%$. (Fig.\ref{Si_corr}c).

\subsubsection{Correction for beam profile $P$.}
\label{Si_B1}

The sensitivity of dynamical electron diffraction to thickness is apparent in the $R_1$ calculation for all integrated intensities (see \ref{Dyn_t}).  It is also very important for the fine structure of rocking curves of strongly dynamical interactions, which show fringes that change in size and number as a function of crystal thickness.  It has long been known that crystal thickness can be measured to an accuracy of a nm or better using these features in CBED patterns.\citet{kelly,allen}  The fine structure of strongly dynamical rocking curves thus gives another way to measure crystal thickness, which should match the minimum $R_1$ for all reflections.  These fringes can be seen clearly for the 311 reflection in Fig.~\ref{Si_rocks}.  However, the features in Fig.~\ref{Si_rocks}f are noticeably sharper than the experimental rocking curve Fig.~\ref{Si_rocks}b and this is due to the angular range of the incident electron beam.  Additionally, in the optimised orientation Fig.~\ref{Si_intersections}b the yellow dots do not lie precisely on a straight horizontal line.  Both of these effects may be explained if the intersections are not points, but have a finite size.  

\begin{figure}
    \caption{a) The direct beam profile, obtained by averaging all frames in the Si cRED data and a Lorentzian fit.  b) $R_1$ calculation after optimising geometry and correcting for slew rate and beam profile (specimen thickness 185 nm). $R_1=8.9\%$. c) $R_1$ for the kinematic model $K$, initial dynamical model $D$, with geometry optimisation $G$, corrections for slew rate $S$ and beam profile $P$, and refinement of Debye-Waller factor $DWF$.}
    \label{Si_beam}
    \centering \includegraphics[width=0.9\columnwidth]{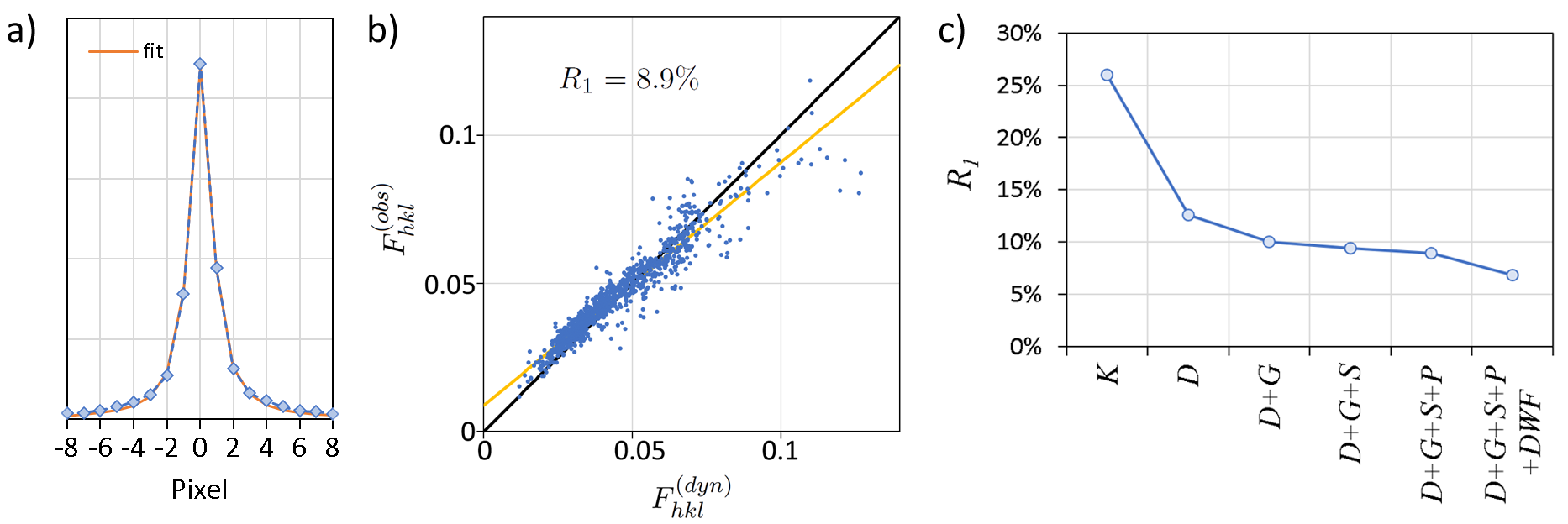}
\end{figure}

The incident electron beam is not a perfect plane wave because the crossover produced by the final condenser lens, which acts as an effective illumination source, is of finite size.  We may approximate the angular profile of the incident beam by the intensity profile of the direct transmitted beam averaged through all frames, which is shown in Fig.~\ref{Si_beam}a together with a fit to a Lorentzian profile.  The fit is excellent and gives the FWHM of the direct beam to be 0.037 $\text{\AA}^{-1}$, or 0.47 mrad (97 arc sec).  Applying this beam profile as a convolution to the simulated LACBED patterns allows rocking curves to be extracted that are a good approximation to experiment, as shown in Figs.~\ref{Si_rocks}g and ~\ref{Si_rocks}h.  Integrated intensities obtained after applying this final correction gives $R_1=8.9\%$ (Fig.~\ref{Si_beam}b).  The largest scatter from the $R_1=0$ line is now found in the highest intensity reflections.

\begin{figure}
    \caption{a) $R_1$ as a function of Debye-Waller factor $B$.  b) Optimised $R_1$ calculation for $B=0.33$, giving a best $R_1=6.8\%$ at a specimen thickness of 185 nm.}
    \label{Si_DWF}
    \centering \includegraphics[width=0.9\columnwidth]{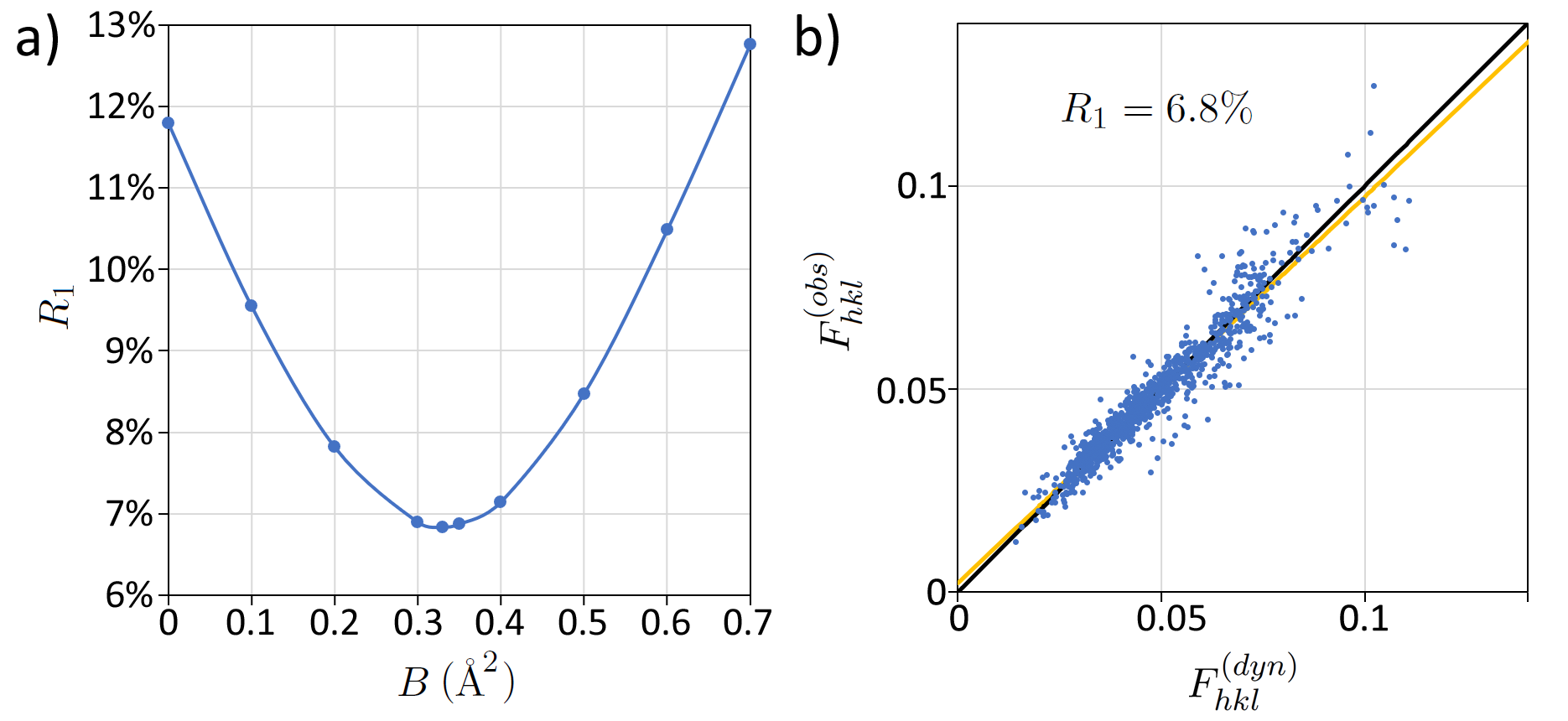}
\end{figure}

Having applied all relevant corrections to the experimental data and optimised the simulation, we are finally in a position to perform a structural refinement.  In silicon there is only one free parameter -- the Debye-Waller factor $B$.  Fig.~\ref{Si_DWF} shows the variation of $R_1$ with $B$, with a final $R_1$ of 6.8\% and a well-defined best-fit at $B=0.32 \text{\AA}^2$.  This is rather lower than the X-ray value of $B=0.54 \text{\AA}^2$ \citet{tobensetal} and there is still noticeable scatter in the highest intensity measurements.  These differences may be due to difficulties in background subtraction, which result from a limited number of intensity measurements in the $PETS$ output (see section \ref{background})

In summary for this silicon cRED data, we have demonstrated that the poor $R_1$ obtained using a kinematical diffraction model is not found when using a dynamical model with appropriate optimisation and correction factors (Fig.~\ref{Si_beam}c).  
The final result ($R_1<7\%$) approaches the values deemed acceptable for X-rays. 
The importance of careful correction is very clear from the observation that the improvement in $R_1$ due to refinement of the temperature factor (Fig.~\ref{Si_DWF}) is of similar magnitude to improvements that result from optimisation of geometry (Fig.~\ref{Si_beam}c).  
Nevertheless, dynamical ED intensities are far more sensitive to structure (i.e. atomic coordinates) than temperature factors \citet{beany2021}, which is reassuring for structure solution and consistent with the growing number of structures solved by ED. Further improvements may be possible with improved background subtraction, which may be incorrect for high intensity reflections in this particular example due to a lack of data in the rocking curves.

\section{Discussion and conclusions}

In this work, we have established a protocol for dynamical modelling of fine-sliced cRED data, taking account of the corrections that should be applied in the case of electron diffraction, equivalent to those applied to X-ray data.  These corrections rely on the ability to extract rocking curves from experimental data, i.e. having a large number of frames collected at small angular increments.  We expect that this approach will become widespread as detector technology continues to improve.  Our results show that dynamical modelling of cRED data has a very significant impact on the quality of fit, reducing $R_1$ by almost 20\% in the silicon example chosen here (similar improvements were seen by \citet{Palatinus2013} for silicon). Nevertheless, this is a particularly simple material with very high perfection and many systematic absences. Equivalent improvements may not be found for more interesting (complex) materials, particularly if they have poorer crystallinity, strong inelastic scattering as seen in organic materials \citet{Latychevskaia:lu5004}, and are not parallel sided lamellae.

Several improvements can be made from this first attempt at dynamical modelling, not least the need for significant computing resources.  Each set of 70 simulations calculated $5.6 \times 10^6$ incident beam orientations, producing a $400 \times 400$ pixel LACBED pattern for each of the 962 diffracted beams.  The large range of reciprocal space covered in each LACBED pattern, together with knowledge of which frame each reflection was seen experimentally, allowed precise correction of the crystal orientation but the area of reciprocal space covered could be reduced by a factor of $>20$ if this optimisation were performed by direct calculation of the positions of the different Bragg conditions.  To allow rocking curves to be captured fully, each simulation overlapped with the next, meaning that every incident beam orientation was simulated twice, something easily avoided if a single simulation (e.g. $20 \times 14000$ pixels) is calculated instead. Rocking curves and integrated intensities could be output directly, rather than extracted from these simulated data using python scripts.  In $Felix$, the Bloch wave calculation is optimised by careful choice of the diffracted beams included \citet{Zuo1995, CHUVILIN200573}, but the time required remains $\propto N^3$, where $N$ is the number of beams \citet{YANG201773}. No attempt was made to optimise this parameter and all simulations were run with $N=200$ from a beam pool of $\sim800$ in each simulation.  If all such improvements were to be implemented it seems reasonable to expect a full set of $|F_{hkl}^{(dyn)}|^2$ to be obtained in seconds, which would then allow dynamical refinement of crystal structure in reasonable times.

Improvements in simulation fidelity are also possible.  The simple Bloch-wave calculation used here assumes that the surface normal is parallel to the incident beam direction, i.e. it does not properly account for continuity of the electron wave function at the entrance and exit surfaces of a tilted crystal.  This is obviously not correct for a specimen tilted by up to $70^\circ$. Furthermore, as the crystal rotates, the thickness of material transited by the electron beam changes.  Interestingly, we found that simulated rocking curves for a single thickness gave a good match to experiment across the full dataset (section \ref{Dyn_r}), and that this thickness agreed with the minimum $R_1$, obtained from integrated intensities.  Furthermore,  incorporating a change in crystal thickness $t$ corresponding to that expected for a parallel-sided slab ($t \propto 1/\text{cos}~\alpha$) gave a worse result than using a single thickness for the complete data set. Presumably these aspects are linked and a more correct model would yield further improvements in $R_1$.  Finally, some experimental rocking curves suffered from poor background correction (section~\ref{background}), although it is unclear for these very strong reflections how much diffuse scattering is present.  More careful extraction of integrated intensities may also improve $R_1$.

In conclusion, dynamical modelling of cRED data has a significant impact on the quality of fit.  If all correction factors are accounted for properly, it seems probable that fit metrics for electron diffraction will equal those of X-ray diffraction.  This is encouraging for future development and application of 3D-ED techniques to structural solution in a wide range of applications.

\section{Acknowledgments}
We would like to thank Dr. Luk{\'{a}}{\v{s}} Palatinus for assistance in using $PETS$ for data processing and to extract rocking curve data.  Data collection was aided by Yani Carter and Will Roberts.  We acknowledge the University of Warwick Research Technology Platforms for Electron Microscopy and Scientific Computing for use of facilities.
AC would like to thank the University of Warwick for PhD funding as part of the Warwick Centre for Doctoral Training in Analytical Science.

\pagebreak
\bibliographystyle{unsrtnat}
\bibliography{references}

\clearpage
\renewcommand{\thesection}{S\arabic{section}}

\def\thefigure{S\arabic{figure}}
\setcounter{figure}{0}
\setcounter{section}{0}
\def\thetable{S\arabic{table}}
\def\theequation{S\arabic{equation}}

\setcounter{figure}{0}
\setcounter{section}{1}
\setcounter{subsection}{0}
\setcounter{table}{0}

\title{\textit{Supplemental Material for}\\%
Modelling fine-sliced three dimensional electron diffraction data with dynamical Bloch-wave simulations}
\maketitle
\label{SuppInfo}

\subsection{Experimental rocking curves and background subtraction.}
\label{background}

\begin{figure}
    \caption{Background subtraction for four rocking curves in the Si cRED data set}
    \label{Backgr}
    \centering \includegraphics[width=1.0\columnwidth]{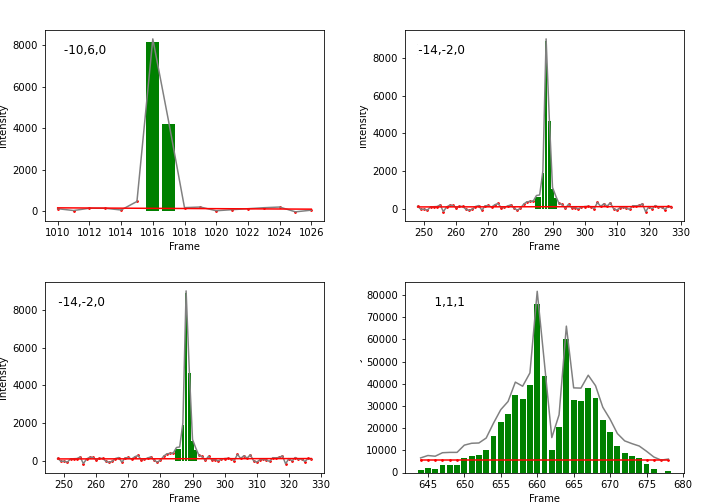}
\end{figure}

Four examples of rocking curves from the silicon cRED data produced by PETS, and subsequent background subtraction, are shown in Fig.~\ref{Backgr}.  The red line is a linear least-squares fit to data points outside the peak. The raw integrated intensity is the sum of the green bars.  For some strong reflections, such as the 111 peak, the data points in the rocking curve may not extend sufficiently to allow a good fit to the background, leading to a systematic underestimation of integrated intensity.

Integrated intensities useful for structure solution and computation of $R_1$s are obtained by applying Lorentz scaling to each raw integrated intensity.

\newpage

\subsection{Direct beam intensity.}
\label{Direct lines}

\begin{figure}
    \caption{The relative direct beam intensity, obtained by cropping the stack of $n=1389$ frames to just the direct beam, producing an average of the beam profile by summing all frames and dividing by $n$, and then dividing each image in the stack by this average.  The stack of frames is then re-sliced to view from the side.  Deficits in the direct beam, caused by each Bragg condition that is passed through as the goniometer rotates, are visible as dark lines.}
    \label{000b}
    \centering \includegraphics[width=0.65\columnwidth]{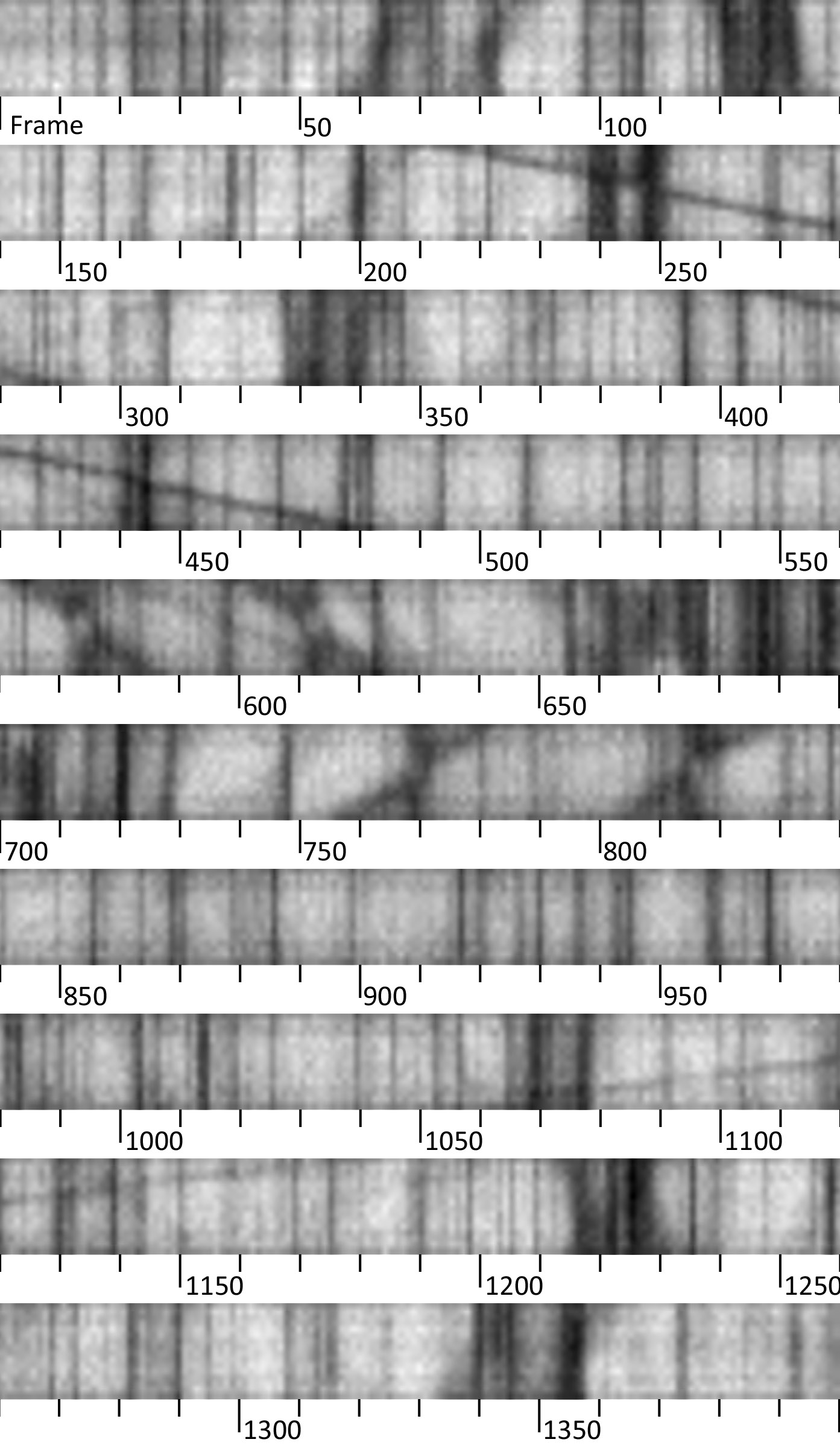}
\end{figure}

\newpage
\subsection{$R_1$ calculation and sensitivity to $B$ in the kinematic model.}
\label{Rfac_eqns}

\begin{figure}
    \caption{$R_1$ calculation as a function of $B$ in the kinematic model.  The lowest $R$ is found at $B=0$.}
    \label{Kin_RvsB}
    \centering \includegraphics[width=1.0\columnwidth]{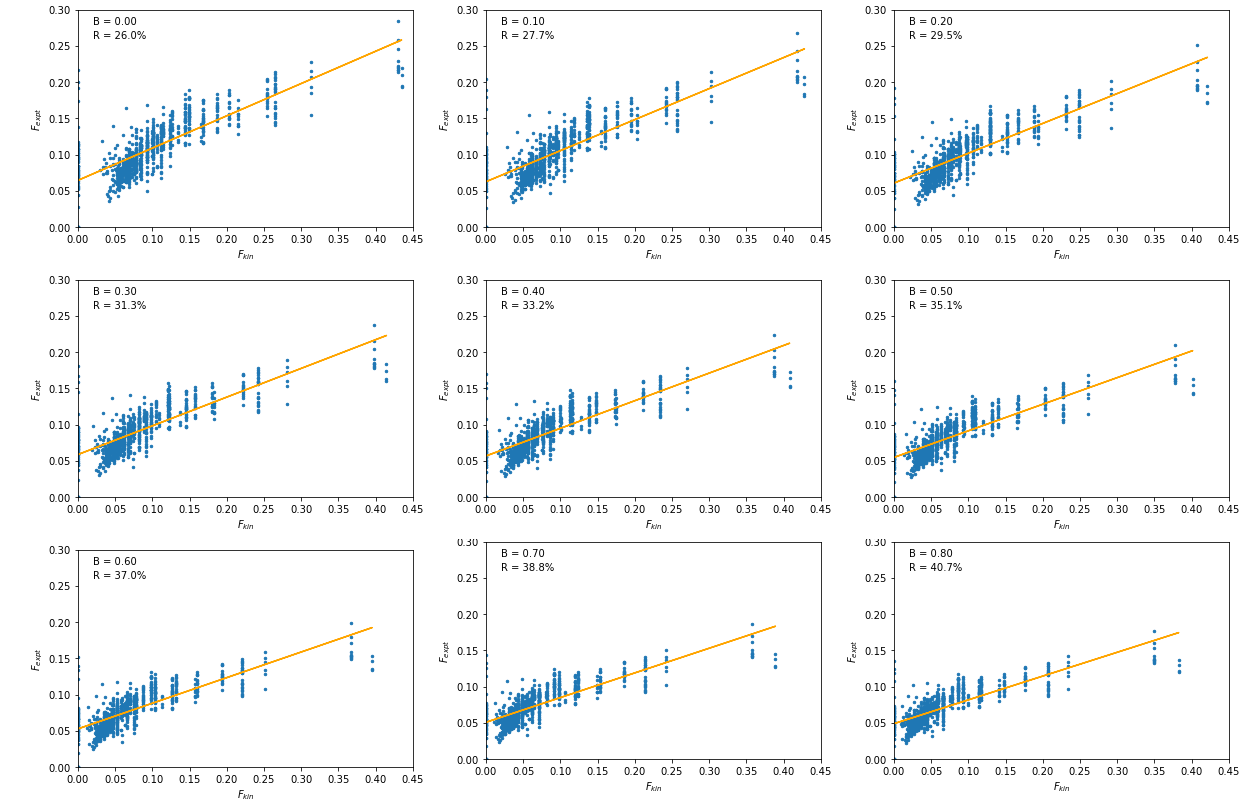}
\end{figure}

In all $R$-factor calculations, we normalise calculated intensities by dividing by $I_{000}=F_{000}F^{*}_{000}$, i.e. the direct beam has an intensity of unity in the absence of any Bragg diffraction.  We take the square root of the experimental Lorentz-scaled integrated intensities $H^{(obs)}_{hkl} = (I^{(obs)}_{hkl})^{0.5}$ and scale them to match calculated structure factors $F^{(calc)}_{hkl}$: 
\begin{equation}
  F_{hkl}^{(obs)} = H_{hkl}^{(obs)} \sum{F_{hkl}^{(kin)} H_{hkl}^{(obs)}}/\sum{H_{hkl}^{(obs)} H_{hkl}^{(obs)}}
  \label{ExpScale}
\end{equation}
where the sum is taken over all observed reflections. The $R_1$ is the obtained from
\begin{equation}
  R_1 = \frac{\left(\sum F_{hkl}^{(obs)} - F_{hkl}^{(calc)} \right)}{\sum F_{hkl}^{(obs)}}, \label{Rfac}
\end{equation}
where the sum is again taken over all observed reflections.  The quality of fit can be seen by plotting $F_{hkl}^{(obs)}$ against $F_{hkl}^{(calc)}$ as shown in Fig.~\ref{Kin_RvsB}.

\newpage
\subsection{Dynamical $R$-factors as a function of thickness.}
\label{Dyn_t}

Since diffracted intensities change significantly as a function of specimen thickness in a dynamical model, each specimen thickness has a different $R_1$.  This is illustrated below in Figs.~\ref{Dyn_Rvst} and \ref{R_vs_t} for Bloch-wave simulations corresponding to the nominal beam path in Fig.~\ref{Si_track}.

\begin{figure}
    \caption{$R_1$ calculation as a function of specimen thickness $t$ in the dynamic model, with the nominal beam path given by $PETS$}
    \label{Dyn_Rvst}
    \centering \includegraphics[width=1.0\columnwidth]{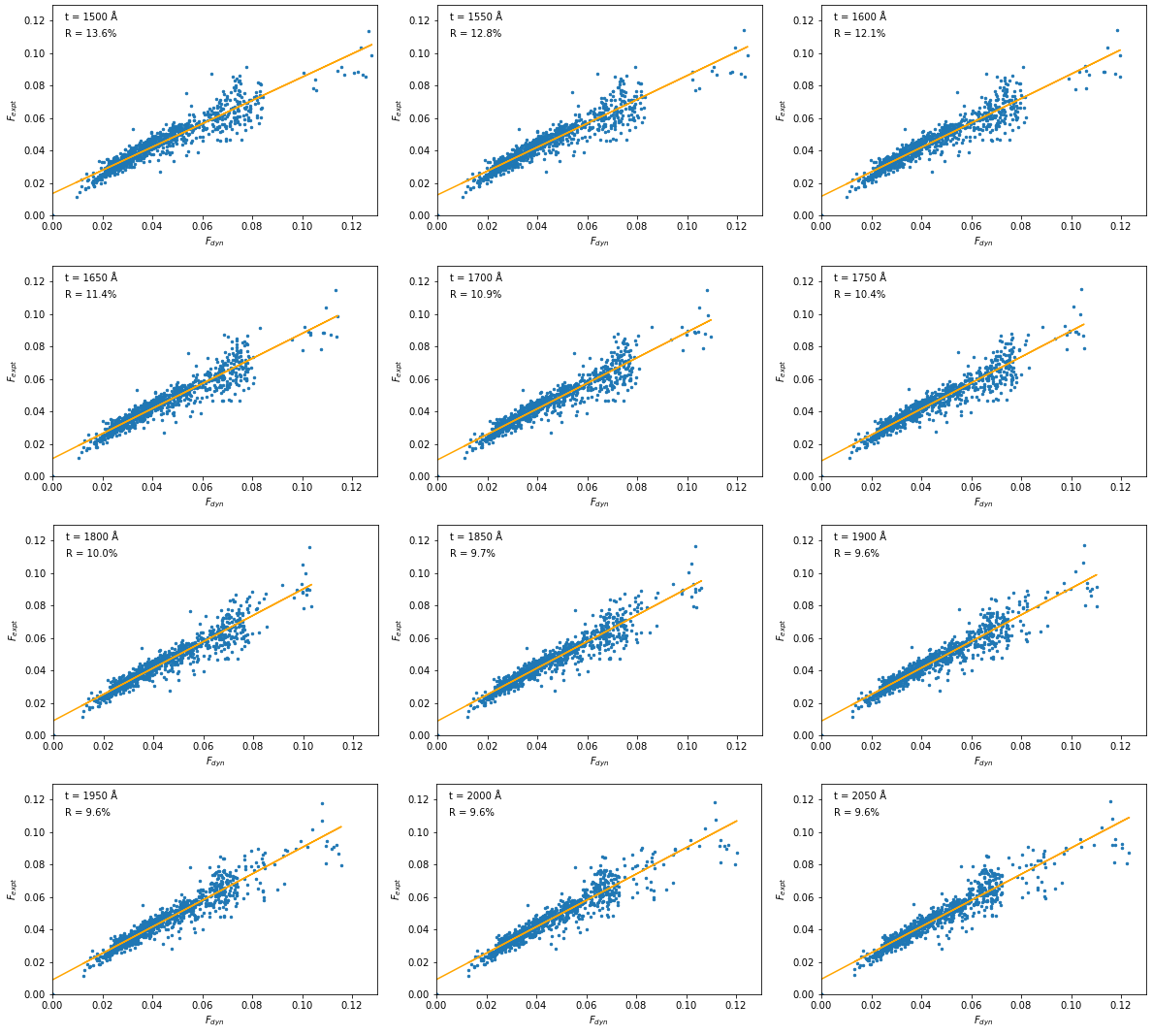}
\end{figure}

\begin{figure}
    \caption{$R_1$ as a function of specimen thickness $t$ in the dynamic model, with an optimised beam path and beam profile convolution.}
    \label{R_vs_t}
    \centering \includegraphics[width=0.5\columnwidth]{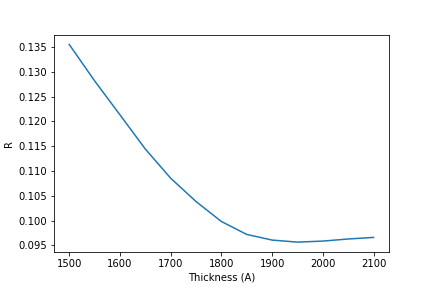}
\end{figure}

\newpage
\subsection{Dynamical rocking curves as a function of thickness.}
\label{Dyn_r}

The fine structure of rocking curves obtained from strongly dynamical reflections is very sensitive to specimen thickness, as exemplified here by Si 311.  The simulated rocking curves have been convoluted with the experimentally measured beam profile.

\begin{figure}
    \caption{Comparison of the experimental Si 311 rocking curve (blue) and simulations at a variety of thicknesses (orange)}
    \label{Rocking Curves v Thickness}
    \centering \includegraphics[width=1.0\columnwidth]{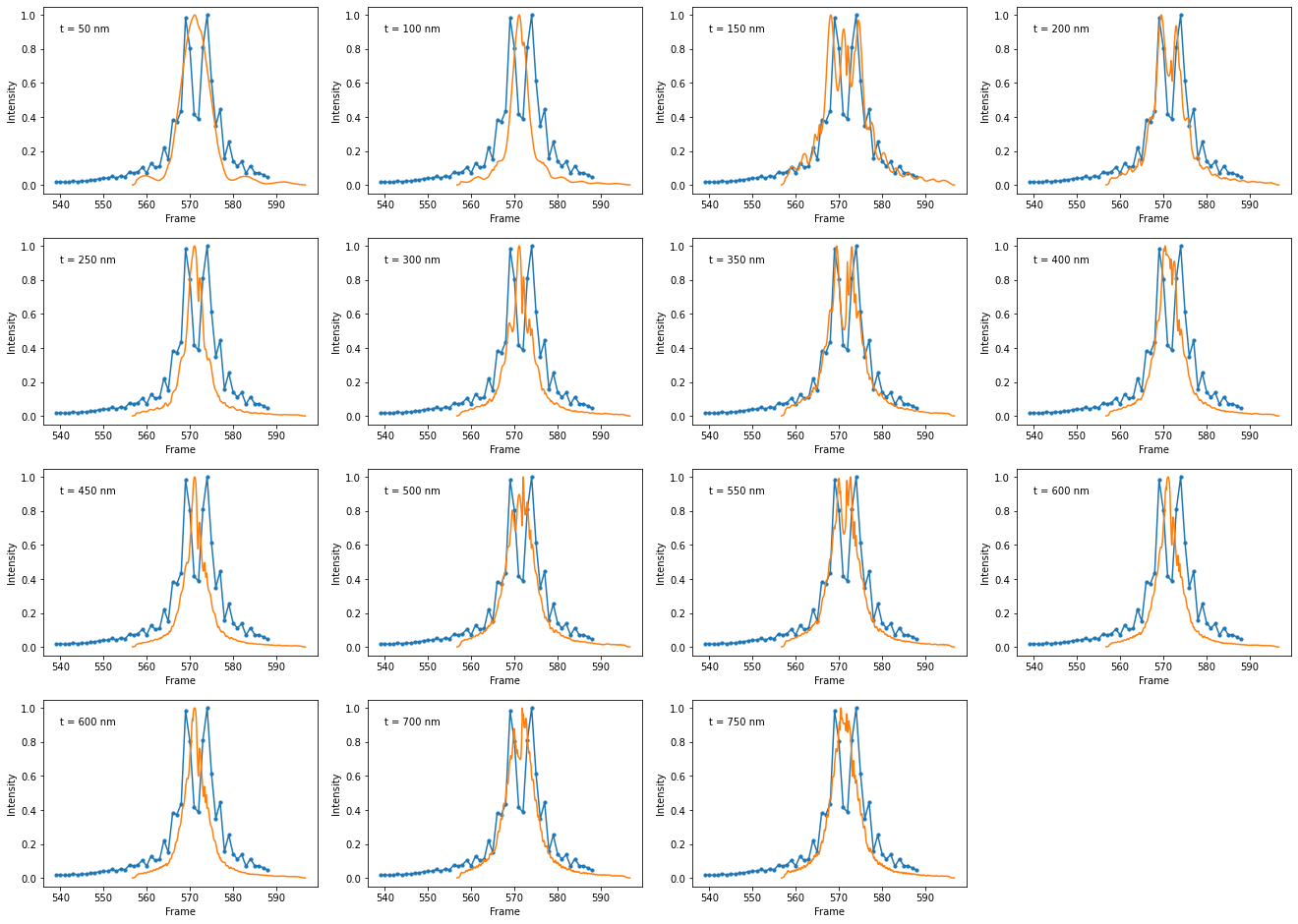}
\end{figure}

\newpage
\subsection{$R$-factor as a function of $B$ for optimised simulations.}
\label{Dyn_B}

$R_1$ is sensitive to thermal vibrations of atoms in the dynamical model as shown in the plots of $F_{hkl}^{(obs)}$ against $F_{hkl}^{(dyn)}$ for different values of the Debye-Waller factor $B$.  The systematic underestimation of strong reflections (\ref{background}) may lead to an underestimation of $B$.

\begin{figure}
    \caption{$R_1$ calculations as a function of $B$ for an optimised dynamical simulation as shown in Fig.~\ref{Si_DWF}.  The lowest $R$ is found at $B=0.33$}
    \label{Dyn_RvsB}
    \centering \includegraphics[width=1.0\columnwidth]{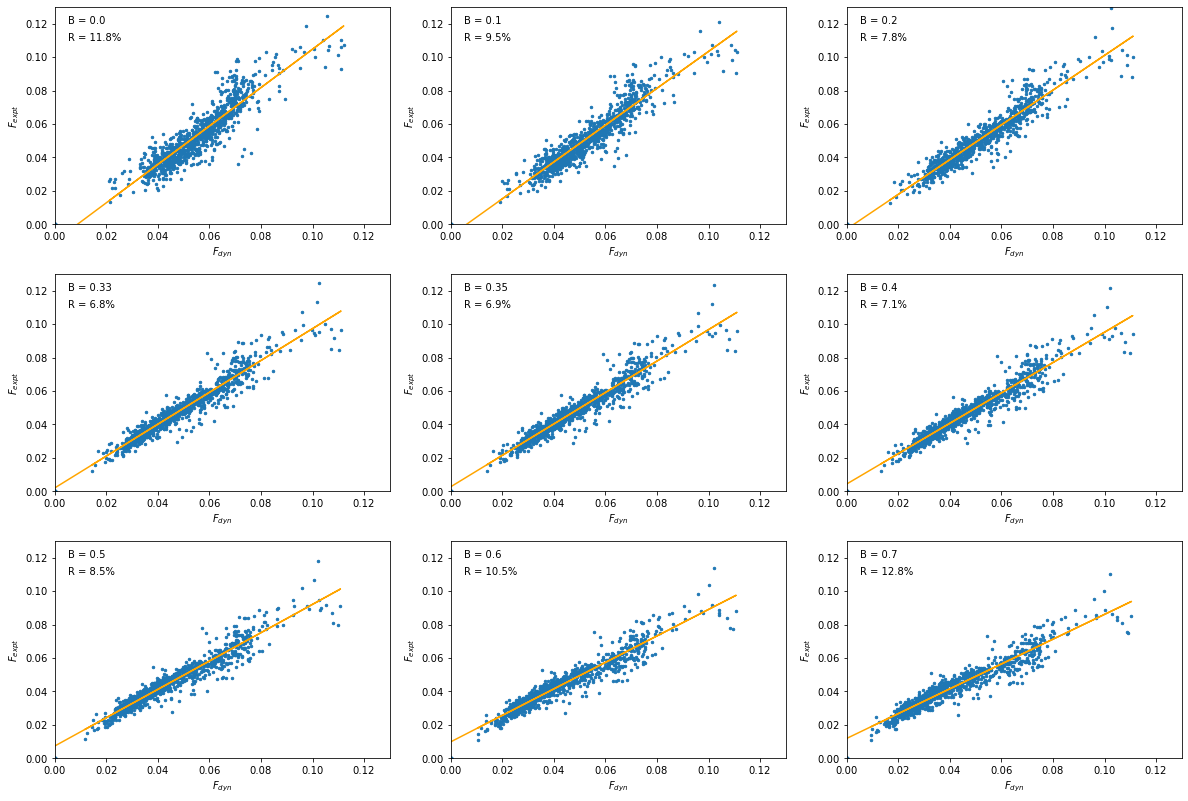}
\end{figure}

\end{document}